\documentclass{iaus}
\usepackage{graphicx}

\title[Two distributions shedding light on supernova Ia progenitors]
{Two distributions shedding light on supernova Ia progenitors: delay times and G-dwarf metallicities}

\author[N. Mennekens, D. Vanbeveren, J.P. De Greve \& E. De Donder]
{Nicki Mennekens$^1$, Dany Vanbeveren$^{1,2}$, Jean-Pierre De Greve$^1$ \and Erwin De Donder$^{1,3}$}

\affiliation{$^1$Astrophysical Institute, Vrije Universiteit Brussel, Pleinlaan 2, 1050 Brussels, Belgium \\
email: {\tt nmenneke@vub.ac.be} \\[\affilskip]
$^2$Groep T - Leuven Engineering College, K.U.Leuven Association, Andreas Vesaliusstraat 13, 3000 Leuven, Belgium \\[\affilskip]
$^3$Belgian Institute for Space Aeronomy (BIRA-IASB), Ringlaan 3, 1180 Brussels, Belgium}

\pubyear{2011}
\volume{281}
\pagerange{1--4}
\jname{Binary Paths to Type Ia Supernovae Explosions}
\editors{R. Di Stefano \& M. Orio, eds.}
\begin{document}

\maketitle

\begin{abstract}
Using a population number synthesis code with detailed binary evolution, we calculate the distribution of the number of type Ia supernovae as a function of time after starburst. This is done for both main progenitor scenarios (single degenerate and double degenerate), but also with various evolutionary assumptions (such as mass transfer efficiency, angular momentum loss, and common envelope description). The comparison of these theoretically predicted delay time distributions with observations in elliptical galaxies then allows to constrain the evolutionary scenarios and parameters. From the morphological shape of the distributions, we conclude that all supernovae Ia cannot be produced through the single degenerate scenario alone, with the best match being obtained when both scenarios contribute. Within the double degenerate scenario, most systems go through a phase of quasi-conservative, stable Roche lobe overflow. We propose stellar rotation as a possible solution for the underestimation of the observed absolute number of events, as is the case in many theoretical population synthesis studies. A brief comparison with these other studies is made, showing good correspondence under the nontrivial condition of equivalent assumptions. We also investigate the influence of different supernova Ia progenitors and evolutionary parameters on the theoretical distribution of the iron abundance of G-type dwarfs in the Galactic disk. These stars are good indicators of the entire chemical history of the Galaxy, and their predicted metallicity distribution can also be compared to the observational ones. This again limits the number of acceptable combinations of assumptions. Supporting previous results, the best correspondence is found in the case where both the single and double degenerate scenario contribute.
\end{abstract}

\firstsection
\section{Introduction}

To distinguish between the single (SD; accretion by a white dwarf (WD) from a late main sequence (MS) or red giant (RG) companion) and double (DD; the spiral-in and merger of two WDs due to gravitational wave radiation emission) degenerate scenario for type Ia supernova (SN Ia) formation, we use an updated version of the population number synthesis (PNS) code by \cite[De Donder \& Vanbeveren (2004)]{dedonder2004}. This code contains detailed binary evolution, without the use of analytical formalisms. SD progenitors are assumed to be as given by \cite[Hachisu et al. (2008)]{hachisu2008}, including the mass stripping effect with its strength parameter $c_1$ which can lie between 0 (turning the effect off) and 10. For the DD scenario, it is either assumed that every WD merger exceeding 1.4 $M_{\odot}$ results in such an event, or that it is additionally required to be a merger of two C-O WDs. Obviously, these assumptions result in different absolute SN Ia rates, however, the observed trends and thus the conclusions are not affected. A parameter study is undertaken, first and foremost concerning the fraction $\beta$ of Roche lobe overflow (RLOF) material that is being accepted by the accretor star. If $\beta<1$, indicating non-conservative RLOF, it is assumed that the lost matter leaves the system through a circumbinary disk, therefore taking with it the specific angular momentum of $L_2$. If a common envelope (CE) phase is encountered, it is treated with the $\alpha$-formalism by \cite[Webbink (1984)]{webbink1984}.

There are two typical channels that can lead to a DD SN Ia: a system can either evolve through a stable RLOF followed by a CE phase (typically resulting in a delay time of a few Gyr) or through two successive CE phases (with a typical delay of only a few tens or hundreds of Myr). It is important to remark however that the former channel only works if the RLOF phase is assumed to be (quasi-)conservative, otherwise systems will already merge during this phase and thus not result in a SN Ia. As in most other PNS studies, we find an absolute SN Ia rate that lies too low compared to observations, in our case by a factor of 2-3 at 11 Gyr. While this may be partially due to errors in the conversion factor between observational (SNuK) and theoretical (SNuM) units (see also the contribution by Pritchet), we present a suggestion for a more physical solution to this problem. There are indications that stars born in binaries rotate faster on the ZAMS than normal single stars. If this is so, they will also have heavier MS convective cores, which will (even if the stars are synchronized later on) result in heavier WD remnants and thus more WD mergers exceeding 1.4 $M_{\odot}$.

\section{Delay time distribution}

\begin{figure}[b]
\begin{center}
 \includegraphics[width=3.4in]{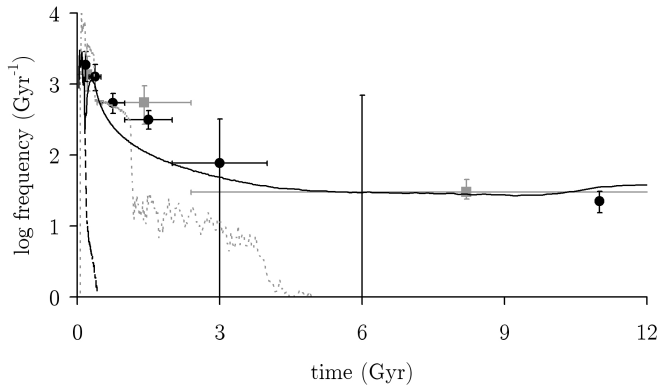} 
 \caption{DD DTDs for $\beta=1$ (solid black) and $\beta=0$ (dashed black), as well as SD DTD (dotted gray). Observations by \cite[Totani et al. (2008)]{totani2008} (black circles) and \cite[Maoz et al. (2011)]{maoz2011} (gray squares).}
   \label{fig1}
\end{center}
\end{figure}

Figure 1 shows the theoretically predicted SN Ia delay time distribution (DTD) for different scenarios and values of $\beta$, as well as two observed DTDs. It is obvious that the SD DTD (shown for $c_1=3$) drops away too fast and too soon in order to keep matching the observations at later times. The DD DTD is able to match the observations, but only in the case of $\beta=1$. This implies not only that $\beta$ indeed needs to lie close to 1, but also that those systems which no longer result in a SN Ia if $\beta=0$ (about 80\% of the total) have gone through such a quasi-conservative RLOF phase, and not two successive CE phases. Most of the SD events issue from the WD+MS channel, not WD+RG. The best match between the observed and predicted DTD is obtained when we assume that both the SD and DD channel contribute, the former with a slightly less high value ($c_1=1$) for the mass stripping parameter than is standard.

A note on the differences in models obtained by different groups performing PNS studies (see also the contributions by Toonen, Nelemans and Claeys). A comparison study is going on to look for the causes of differences in the predictions. It should finally result in an explanatory paper. The most important discrepancies are being caused by differences in assumptions, mainly about mass and angular momentum loss and the treatment of CE phases. Homogenizing the assumptions as much as the codes technically allow results in converging models. Some (more minor) discrepancies remain. This seems inevitable given the different treatment of the ``single star tracks'', i.e. the stellar evolution of individual binary components. For example, the use of the \cite[Hurley et al. (2002)]{hurley2002} prescription, versus using full binary evolution.

\section{G-dwarf metallicity distribution}

\begin{figure}[b]
\begin{center}
 \includegraphics[width=3.4in]{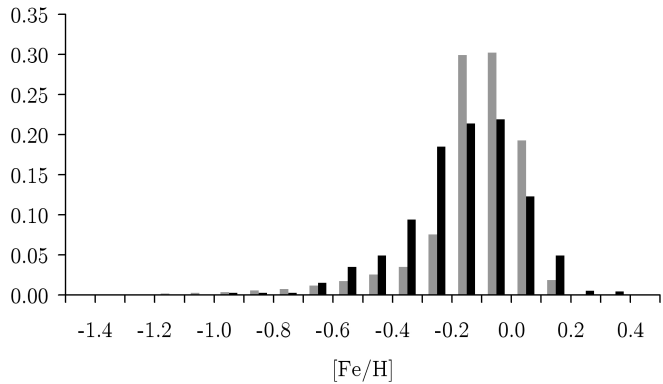} 
 \caption{G-dwarf metallicity distribution with two-infall model for SD+DD (gray histogram). Spherical solar neighborhood observations by \cite[Holmberg et al. (2007)]{holmberg2007} (black histogram).}
   \label{fig2}
\end{center}
\end{figure}

\begin{figure}[b]
\begin{center}
 \includegraphics[width=3.4in]{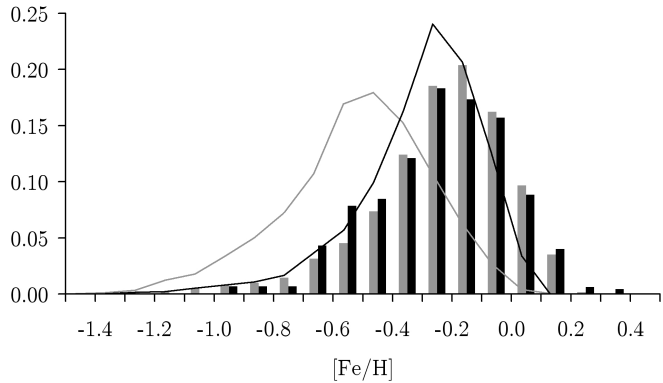} 
 \caption{G-dwarf metallicity distribution with constant SFR for SD+DD (gray histogram), as well as for SD (gray line) and DD (black line) only. Cylindrical solar neighborhood observations by \cite[Holmberg et al. (2007)]{holmberg2007} (black histogram).}
   \label{fig3}
\end{center}
\end{figure}

A second distribution which allows to constrain SN Ia progenitors is the G-dwarf metallicity distribution. Because of their very long lifetime, G-type dwarfs in the Galactic disk are excellent indicators of the entire chemical history of this region. The metallicity ([Fe/H]) distribution of these stars is critically affected by the SN Ia rate throughout the Galaxy's history, and thus also by the progenitor assumptions. The thus predicted distributions can be compared to observed distributions for a cylindrical solar neighborhood, in this case those by \cite[Holmberg et al. (2007)]{holmberg2007}. We use an updated version of the chemical evolution model used by \cite[De Donder \& Vanbeveren (2004)]{dedonder2004} for a similar study. Other studies have been performed in the meantime by e.g. \cite[Greggio et al. (2008)]{greggio2008} and \cite[Matteucci et al. (2009)]{matteucci2009} (both not with a full PNS model but with various adopted DTDs), as well as very recently by \cite[Kobayashi \& Nakasato (2011)]{kobayashi2011} (for SD progenitors only). We use a binary fraction (the fraction of stars with a stellar companion) of 70\%, which is required to be that large in order to yield sufficiently high SN Ia rates. For the galaxy formation, we assume either the two-infall model or a constant star formation rate (SFR).

In case of the SFR obtained from the two-infall model, and for contribution by both the SD (with $c_1=1$) and DD channel, we find a metallicity distribution which peaks at approximately the right location, however it is far too much peaked to match the morphological shape of the observed one. Figure 2 illustrates that it shows a slightly better resemblance to the \cite[Holmberg et al. (2007)]{holmberg2007} distribution for a spherical solar neighborhood as is used for comparison by \cite[Kobayashi \& Nakasato (2011)]{kobayashi2011}. However, since in our code a cylindrical model is used, we should indeed compare to such observations.

The metallicity distribution which is found when a constant SFR (with total star formation equal to the previous case) is assumed, is shown in Fig. 3. In addition to cylindrical model observations, this figure shows the histogram obtained with the combined SD (with $c_1=1$) + DD model, as well as the models for only SD (with $c_1=3$) and DD respectively. It shows that the combined model produces a very good match, while either scenario alone (especially the SD, even with high $c_1$) is unable to produce sufficiently high metallicity. Therefore, this study with its updated and internally PNS-computed SN Ia yields supports previous results favoring the SD + DD model. This combined model also satisfactorily reproduces the observed [C/Fe] and [O/Fe] vs. [Fe/H] relations.

\section{Conclusions}

We find that both the delay time (see also \cite[Mennekens et al. (2010)]{mennekens2010}) and G-dwarf metallicity distribution point toward a significant contribution by both the single and double degenerate scenario, the latter mainly through systems that have undergone a quasi-conservative Roche lobe overflow followed by a common envelope phase. The critical dependence of both distributions on certain binary evolutionary processes, exactly those processes which are described in population codes by still uncertain parameters, is a way to find out more about these processes and thus further constrain these parameters.

\end{document}